\documentstyle[10pt]{article}
\makeatletter
\@addtoreset{equation}{section}
\makeatother

\font\ninerm=cmr9
\font\nineit=cmti9

\font\eightrm=cmr8

\newcommand{\ncm}{\newcommand}
\ncm{\rencm}{\renewcommand}
\begin{document}
\newpage
%
\hbox{}
\mbox{}\hspace{0.5cm}June 1992 \hspace{8.0cm} HLRZ-92-31\\
\mbox{} \hspace{10.6cm}  BUTP-92/29\\
\begin{center}
\vspace*{0.5cm}
{\LARGE A Multicanonical Algorithm\\ and the Surface Free
Energy\vspace{0.1cm}\hfill\\in SU(3) Pure Gauge Theory}\\
\vspace*{0.7cm}
{\large B.~Grossmann$^{1}$, M.~L.~Laursen$^{1}$,\\
T.~Trappenberg$^{1,2}$,
        U.-J.~Wiese$^{3,}$\footnote{Supported by Schweizer
Nationalfond} \\}
\vspace*{0.7cm}
{\small
        $\mbox{}^{1}$HLRZ,c/o Kfa Juelich, D-5170 Juelich, Germany\\
        $\mbox{}^{2}$Inst. f. Theor. Physik, RWTH Aachen, D-5100
Aachen, Germany\\
        $\mbox{}^{3}$Inst. f. Theor. Physik, Univ. Bern, CH-3012
Bern, Switzerland.}\\
\vspace*{2cm}
{\bf Abstract}
\end{center}
We present a multicanonical algorithm for the SU(3) pure gauge theory
at the
deconfinement phase transition. We measure the tunneling times for
lattices of
size $L^3\times 2$ for $L=8,\, 10,$ and $12$. In contrast to the
canonical
algorithm the tunneling time increases only moderately with L. Finally,
we
determine the interfacial free energy applying the multicanonical
algorithm.

\setlength{\baselineskip}{1.3\baselineskip}
%
%
%
%
\section{Introduction}
For the analysis of systems with first order phase transitions--as e.g.
quenched
QCD at the deconfinement temperature--surface effects are of crucial
importance.
They determine the physical dynamics and lead to supercritical
slowing down in computer simulations at the critical point.
Configurations
which include an interface are
exponentially suppressed by their free energy which results in an
exponential increase
of the tunneling time. Recently the multicanonical algorithm was
proposed in
order to overcome this supercritical slowing
down \cite{ber91}.
 Here we present a variant of this method for
the SU(3) gauge theory with the Wilson action. We have used lattices
of size
$L^3\times L_T$ with $L_T=2$ and $L=8,\, 10,$ and $12$ for which
there is a
strong first order phase transition. We achieve a considerable
reduction in the tunneling time with only a modest increase
in computer time.
We apply this algorithm to the determination of the
interfacial free energy
\begin{equation}
F_{cd} = \alpha_{cd} A
\end{equation}
of a confinement-deconfinement interface of area $A$, $\alpha_{cd}$
being its
surface tension. For this we made use of Binder's histogram method
which in
contrast to the methods used in \cite{hua90,kaj90} generates interfaces
dynamically rather than by introducing artificial external parameters.
In that respect, this approach is similar to transfer matrix
methods \cite{gro91}.

\section{Monte Carlo Simulations at a First Order Phase Transition}
As mentioned above, MC simulations close to first order phase
transitions suffer
from supercritical slowing down, i.e. the tunneling time $\tau_L$ for
a system
of volume $V=L^3$ is expected to be proportional to the inverse of the
probability of a system with  interfaces (actually there will be two
interfaces
because of the boundary conditions),
\begin{equation}
\tau_L\propto \exp (2 L^{2}\frac{\alpha}{T} )
\end{equation}
and therefore diverges exponentially with the area $A=L^{2}$ of an
interface.

In order to overcome this problem, the multicanonical algorithm does
not sample
the configurations with the canonical Boltzmann weight
\begin{equation}
{\cal P}_L^{can}(S) \propto \exp (\beta S),
\end{equation}
where $S=1/3\,\sum_\Box\,tr\,U_\Box$ is the Wilson action in four
dimensions,
but rather with a modified weight
\begin{equation}
{\cal P}_L^{mc}(S) \propto \exp (\beta_L(S) S+\alpha_L(S)).
\label{eq:boltzmc}
\end{equation}
The coefficients $\alpha_L$ and $\beta_L$ are chosen such that the
probability $P_L$ (not to be confused with the Boltzmann weights) is
increased
for all values of the action in between the two maxima $S_L^{max,1}$
and $S_L^{max,2}$ , as shown schematically in figure~1.
%
Finally data are analyzed from the multicanonical samples by
reweighting with
$\exp (  (\beta - \beta_L(S)) S  -\alpha_L(S))$.

In order to approximate the weights ${\cal P}_L^{mc}$ leading to the
distribution $P_L^{mc}$ of figure~1
, we start from
some good estimate of the canonical distribution $P_L$ (see below) at the
coupling $\beta_L^c$ which corresponds to equal weight in both phases.
This will
lead to different heights of the two peaks.
Then we take a partition
\begin{eqnarray}
 S_L^0 =-6L^3L_t & < & S_L^1\equiv S_L^{max,1}< S_L^2 < \ldots <
S_L^{N/2}\equiv S_L^{min} < \ldots \nonumber\\& \ldots & < S_L^{N-1}
\equiv S_L^{max,2}
< S_L^N=6L^3L_t
\end{eqnarray}
of the interval $-6L^3L_t\le S\le 6L^3L_t$. The coefficients $\alpha_L$
and $\beta_L$
are chosen to be constants $\alpha_L^k$ and $\beta_L^k$ in the intervals
$[ S_L^k,S_L^{k+1} )$ such that $\ln P_L^{mc}$ interpolates the linear
function
\begin{equation}
 \ln P_L^{max,1} + (\ln P_L^{max,2}- \ln P_L^{max,1}) \cdot
   \frac{S-S_L^{max,1}}{S_L^{max,2}-S_l^{max,1}} .
\end{equation}
continuously between the points $S_L^k$ and $S_L^{k+1}$ for $k=1$ to
$N-2$. The
Boltzmann weight is identical to the canonical one in the first and
last interval.
We arrive at
\begin{equation}
 \beta_L^k-\beta = \left\{
 \begin{array}{ll}
      0 & , k=0 \\
      \delta\beta +  \ln\left(\frac{P_L(S_L^k)} {P_L(S_L^{k+1})} \right)
                           / (S_L^{k+1}-S_L^k)
        & ,k=1,\ldots ,N-2 \\
 0 &  ,k=N-1
 \end{array}
                \right.
\end{equation}
where $\delta\beta\equiv \ln\left( \frac{P_L^{max,2}}{P_L^{max,1}}
\right)
                         /(S_L^{max,2} - S_L^{max,1})$, and the
$\alpha_L^k$ are
given by
\begin{equation}
 \alpha_L^{k+1} \equiv \alpha_L^k + (\beta_L^k - \beta_L^{k+1})
S_L^{k+1},\;
                \alpha_L^0 = \delta\beta\cdot S_L^{max,1}.
\end{equation}
The partition $\{S_L^k\}$ which is in principle arbitrary is defined
by demanding
\begin{eqnarray}
 \frac{P_L (S_L^{k+1})}{P_L(S_L^k)} = \left\{
  \begin{array}{ll}
  1/r_1        &,k=1,\ldots,N/2-1 \\
  r_2          &,k=N/2,\ldots,N-2,
  \end{array}
\right.\nonumber\hfill (2.8)
\end{eqnarray}
such that
$r_1^{N/2-1}=P_L^{max,1}/P_L^{min}$ and
$r_2^{N/2-1}=P_L^{max,2}/P_L^{min}$. This
generalizes the formulas given in \cite{ber91} to the case
$P_L^{max,1}\neq
P_L^{max,2}$ by introducing a $\delta\beta\neq 0$. In addition the
specific
choice for $r_1$ and $r_2$  assures that the
probability of $S_L^{min}$ is lifted by the correct amount.

We apply this algorithm to the $SU(3)$ pure gauge theory  at the
deconfinement
phase transition. The multicanonical data sampling was done with a
5-hit
Metropolis algorithm as well as with a Creutz heat bath algorithm
modified
according to equ.~(\ref{eq:boltzmc}). In both cases  three
independent $SU(2)$-subgroups
are updated following the idea of ref.~\cite{cab82}.
Because of the dependence of $\alpha_L$ and
$\beta_L$ on the total action $S$, the update of the active link has
to be done
in scalar mode. However, most of the computation time is needed for the
calculation of the staples surrounding the active link, and this is
still
vectorizable. In addition, for the Creutz
algorithm we profited from the fact that as long as the action stays
within one
interval $[S_L^k,S_L^{k+1})$ there is no dependence of $\alpha_L$ and
$\beta_L$
on $S$. Therefore, if the intervals are not too small  even the update
can be
vectorized partially, since one hardly ever crosses an interval boundary.

 In table \ref{tab:times} the update times for the various algorithms
for one processor of a CRAY-YMP are given for a $10^3\times 2$ lattice
using $N=14$ intervals for ${\cal P}_L^{mc}$. As can be seen the CPU
time per link and
update is
increased by only about $50\%$.
\begin{table}[ht]
\smallskip
  \centerline{\eightrm Table 1. CPU time per link update for canonical
and
   multicanonical}
  \centerline{\eightrm   versions of
   a Metropolis and a Creutz heatbath algorithm.}
\smallskip
  \label{tab:times}
\begin{center}
\begin{tabular}{|c||c|c|c|c|} \hline
   Algorithmus & Metr. & Metr.(mc) & Creutz & Creutz(mc) \\ \hline\hline
   $t_{upd}(\mu sec)$ & 37 & 53 & 31 & 52 \\   \hline
\end{tabular}
\end{center}
\medskip\noindent
\end{table}
 In our simulations we use $C$-periodic boundary conditions in the
three spatial
directions (see \cite{wie91}). This
suppresses  the two deconfined phases with nonvanishing expectation
value of the
imaginary part of the Polyakov line. Then only the confined and one
deconfined phase coexist at the critical temperature $T_c$. For finite
volumes we define
$T_L^c$ (or $\beta_L^c$ resp.) by
demanding equal weight for these two phases \cite{bor92}, i.e.
\begin{equation}
  P_L(S\leq S_L^{min}) \equiv P_L(S\geq S_L^{min}).
\end{equation}

We typically do $50000$ updates for $\beta\approx \beta_L^c$. One sweep
consists
of one Metropolis (resp. Creutz) update step plus four overrelaxation
steps.
For the smaller lattices $L=8$ and $10$ the parameters for the
multicanonical
algorithm were taken from canonical simulations. For $L=12$ where the
canonical
algorithm is no longer efficient due to the increasing tunneling time,
we took
the parameters from an appropriate finite size scaling extrapolation
of the
distributions for the $10^3\times 2$ lattice. In  all  cases we take
$N=14$
intervals.

Figure 2
 shows part of the history of the average Polyakov line as a
function of the number of sweeps for the various algorithms.
{}From the full histories we extract the  tunneling times $\tau_L$,
defined as one
fourth of the number of sweeps needed for a full tunneling cycle.
The   tunneling times $\tau_L$ in table \ref{tab:auto}
indicate that the  increase with $L$ for the standard algorithm is
effectively reduced. Already for $L=10$ the gain in $\tau_L$ is much
larger than the overhead in CPU time per update which is due to scalar
operations.
\begin{table}[ht]
\smallskip
\centerline{\eightrm Table 2. Tunneling times $\tau_L$ for the various
algorithms
            and lattices of size $L^3\times 2$. }
\smallskip
\begin{center}
\begin{tabular}{|c||c|c|c|c|}    \hline
   L & $\beta$ &Metr. & Metr.(mc) & Creutz(mc) \\   \hline\hline
   8  & 5.094 & 300 &  200 & -- \\      \hline
   10 & 5.095 & 1500 & 400 & 400 \\      \hline
   12 & 5.0928 &  --  & 700 & 700  \\      \hline
\end{tabular}
\end{center}
\medskip\noindent
lattices of
  \label{tab:auto}
\end{table}

Thus the multicanonical algorithm provides an  efficient tool for
Monte Carlo studies
of the deconfinement phase transition. As an
application, we present in the next section the determination of
the interfacial
free energy $F_S$.

\section{The Interfacial Free Energy}

The configurations with $S=S_L^{min}$ are expected to contain two
parallel
interfaces of area $L^{2}$ between the confined and the deconfined
phase.
Therefore they should be suppressed by the interfacial free energy
$\alpha L^{2}$.
We expect additional power law corrections due to capillary wave
fluctuations and
translational Goldstone modes of the interfaces, such that
\begin{equation}
 P_L^{min} \propto  L^x\exp\left(-2L^{2}\frac{\alpha}{T}\right)
\end{equation}
(see also ref.\cite{bin81}). The probability distribution should be
flat around
$S_L^{min}$ for non-interacting surfaces due to translations of the
interfaces relative
to each other.

We extract the probability distribution $P_L$ for the plaquette and
the real
part of the Polyakov line from our
multicanonical simulations for $L=8,10,$ and $12$ (see figure 3).
They agree reasonably
with those obtained in ref.~\cite{jan92} for periodic boundary
conditions. However, there
is no plateau for the probabilities around $S_L^{min}$. This is likely
to be due to
interfacial interactions  which will be reduced by using a larger
extension in
one of the spatial directions. This is left for further investigations.

Nevertheless, we consider   the two quantities
\begin{equation}
 F_L^{(1)} \equiv \frac{1}{2L^{2}}
\ln\frac{\overline{P_L}^{max}}{P_L^{min}}
 \end{equation}
and
\begin{equation}
 F_L^{(2)} \equiv -\frac{1}{2L^{2}} \ln P_L^{min}
  \end{equation}
 where $\overline{P_L}^{max}\equiv
\frac{1}{2}(P_L^{max,1}+P_L^{max,2} )$. They
should converge to $\frac{\alpha}{T}$ for $L\rightarrow\infty$ for
non-interacting
surfaces. We plot them as functions  of  $1/L^2$ in figure~4.
 We find a reasonable agreement between the values taken from the
plaquette and Polyakov line distributions.
As expected, the discrepancy between
$F_L^{(1)}$ and $F_L^{(2)}$  decreases for increasing $L$.
Still, compared to the values of $\alpha$ obtained by different methods
\cite{hua90,gro91} the extrapolated result of our data is too small.
Simulations on asymmetric lattices should clarify this point.

In conclusion we have demonstrated the power of the multicanonical
algorithm for SU(3) pure gauge theory. The tunneling time $\tau_L$ is
reduced
considerably while the update time is only increased by less than
$50\%$ compared
to standard
Metropolis or heat bath algorithms. This reduces the computer time
needed for
investigations of the deconfinement phase transition like the one
presented here for
the interfacial free energy. Using cylindrical lattices will probably
reduce the
interactions between the interfaces thereby allowing the extraction of
the interface
tension.

\newpage
\section{Figure Captions}
\begin{itemize}
\item[Figure 1]{Schematic representation of of the probability
distributions $P_L$
(straight line) and $P_L^{mc}$ (dashed line) corresponding to the
canonical
Boltzmann weight
${\cal P}_L^{can}$ and the multicanonical one ${\cal P}_L^{mc}$}.
For $S <
S_L^{max,1}$ and $S> S_L^{max,2}$, $P_L$ and $P_L^{mc}$ (normalized
to their
maximum value) are identical.
\label{fig:pmc}
%
    \item[Figure 2]{Evolution of the average Polyakov line  as a
function of the total
     number of sweeps for a $10^3\times 2$ lattice for (A) a
canonical Metropolis
     algorithm, (B) a multicanonical Metropolis algorithm, and (C) a
multicanonical
     Creutz algorithm. }
     \label{fig:history}
%
    \item[Figure 3]{Distribution for the average plaquette  and the
real part of
                    the Polyakov line
                    for lattice sizes $L=8,\, 10$ and $12$.}
     \label{fig:pdist}
from the average
versus $\ln L/L^2$.
indicated by a diamond,
    \item[Figure 4]{The free energies $F_L^{(1)}$ and $F_L^{(2)}$
from the average
              plaquette  and the real part of the Polyakov line
versus $1/L^2$.
    The value for $\alpha/ T$ from reference \cite{hua90} is indicated
by a diamond,
    the value from reference \cite{kaj90} by a filled square, and
     the value from reference \cite{gro91} by a triangle.}
     \label{fig:fl2}
\end{itemize}
\end{document}